\documentclass[conference]{IEEEtran}
\IEEEoverridecommandlockouts
\usepackage{cite}
\usepackage{amsmath,amssymb,amsfonts}
\usepackage{algorithmic}
\usepackage{graphicx}
\usepackage{textcomp}
\usepackage{xcolor}
\def\BibTeX{{\rm B\kern-.05em{\sc i\kern-.025em b}\kern-.08em
    T\kern-.1667em\lower.7ex\hbox{E}\kern-.125emX}}
    
\usepackage{enumitem}
\usepackage{booktabs}
\usepackage{microtype}

\newcommand{\compresslist}{
  \setlength{\itemsep}{1pt}
  \setlength{\parskip}{0pt}
  \setlength{\parsep}{0pt}
}

\newcommand{\comm}[1]{}

\begin{document}

\title{"I Packed My Bag and in It I Put...": A Taxonomy of Inventory Systems for Virtual Reality Games
\thanks{978-1-6654-3886-5/21/\$31.00 ©2021 IEEE}
} %

\author{\IEEEauthorblockN{1\textsuperscript{st} Sebastian Cmentowski}
\IEEEauthorblockA{\textit{High Performance Computing Group} \\
\textit{University of Duisburg-Essen}\\
Duisburg, Germany \\
sebastian.cmentowski@uni-due.de}
\and
\IEEEauthorblockN{2\textsuperscript{nd} Andrey Krekhov}
\IEEEauthorblockA{\textit{High Performance Computing Group} \\
\textit{University of Duisburg-Essen}\\
Duisburg, Germany \\
andrey.krekhov@uni-due.de}
\and
\IEEEauthorblockN{3\textsuperscript{nd} Jens Kr\"uger}
\IEEEauthorblockA{\textit{High Performance Computing Group} \\
\textit{University of Duisburg-Essen}\\
Duisburg, Germany \\
jens.krueger@uni-due.de}
}

\maketitle

\begin{abstract}
On a journey, a backpack is a perfect place to store and organize the necessary provisions and tools. Similarly, carrying and managing items is a central part of most digital games, providing significant prospects for the player experience. Even though VR games are gradually becoming more mature, most of them still avoid this essential feature. Some of the reasons for this deficit are the additional requirements and challenges that VR imposes on developers to achieve a compelling user experience. We structure the ample design space of VR inventories by analyzing popular VR games and developing a structural taxonomy. We combine our insights with feedback from game developers to identify the essential building blocks and design choices. Finally, we propose meaningful design implications and demonstrate the practical use of our work in action.
\end{abstract}


\section{Introduction}
Inventories are among the most common features in various game genres. Dating back to the beginning of digital games, inventories have evolved from pure item collections storing the players' possessions to sophisticated gameplay features. The range of use cases includes storing and switching items, displaying information, and managing the inventories' contents. Recent games have discovered the inventory as part of the game world and introduced mechanics to make the storage interface more compelling. The game~\textit{Green Hell}~\cite{GameGreenHell} demonstrates how inventories can seamlessly blend into the main gameplay by forcing players to rotate and align their items carefully to fit within a very confined space.

One game platform getting notable attention throughout the last few years is virtual reality (VR). Players use head-mounted displays (HMDs) and tracked controllers to replace their real surroundings with a virtual world. However, one crucial prerequisite to guarantee a compelling and immersive experience is a proper interaction concept. Considering that inventories are the cardinal point of object interaction in many games, they could also provide a compelling addition to the VR experience. Allowing players to carry their found items and building personal storage is a natural addition to this interaction-centered gameplay. Unfortunately, most VR developers still refrain from using inventories and thus fail to reach their game's full potential and profundity.

The causative reasons are plentiful and mostly reside within the additional requirements, such as the inventory's positioning. In contrast to desktop games, VR titles must place the interface in the players' sight without obstructing the surrounding. At the same time, players experience the virtual environment as a substitute for reality, leading to an increased sensitivity for incoherent and unnatural interactions. These obstacles make it challenging to transfer existing non-VR inventories to the virtual world. For instance, an abstract 2D menu works well for desktop games but performs poorly in VR~\cite{handSurvInt}. 

Applying existing research on VR menus to inventories is also not trivial. Most other user interfaces, such as game settings, are designed as abstract overlays prioritizing interaction speed and simplicity. In comparison, inventories are closely tied to the virtual environment and require completely different interactions. For instance, adding an item to the inventory means transferring it from the 3D world to the local storage interface. This transition might even include a resizing or remapping to 2D. In sum, designing storage systems for VR is by no means a trivial task. Paired with a general lack of focused research, these challenges provide a strong motivation for a closer look at VR inventory design.

Our work aims to bridge this gap by forming a structural research foundation, encompassing the status quo, and highlighting interesting research directions. Our main contribution consists of three parts (cf. Figure~\ref{fig:flowchart}). In the first segment, we assess the current state of the art. Therefore, we summarize the relevant related research, collect detailed feedback from active developers and practitioners through semi-structured interviews, and provide an in-depth analysis of current VR games that use inventories as part of their gameplay. In the next part, we combine all three pillars into a condensed framework, which consists of user- and game-related requirements and a comprehensible structural taxonomy summarizing the essential building blocks. As the final step, we demonstrate the practical applicability of our work. We use the presented framework to design three inherently different inventories. This design process is used to discuss the remaining open questions, in particular, the effects and connections between requirements and design choices. This work is meant to build a foundation and inspire future research on this unexplored and multi-faceted topic by raising interesting open questions.

\section{Related Work}\label{RelatedWork}
Despite being one of the most common elements in games, only two closely related works address inventories in VR: Wegner et al.~\cite{wegner2017comparison} compare two concepts for their suitability in serious games, and Cmentowski et al.~\cite{cmentowski2019inventoryLBW} present different inventory designs and establish an early taxonomy. Considering the sparse pool of closely related work, we briefly introduce the most relevant work dealing with VR menus in general. For a detailed overview of menus and interactions in virtual environments, we point to the work by Dachselt et al.~\cite{dachseltMenuSurv}, Kim et al.~\cite{kim2000multimodal}, and Bowman et al.~\cite{bowman20043d}. Unfortunately, the established insights are only partially applicable since inventories differ from most of the researched interfaces. Unlike other menus, such as game settings, inventories should blend into the active gameplay and support a specific set of interactions.

In one of the earliest works on virtual menus, Jacoby et al.~\cite{jacobyVirtualMenu} present seven interaction aspects: \textit{invocation, location, reference frame, cursor, highlighting, selection, and removal}. These terms partially overlap with the design characteristics \textit{placement, selection, representation, and structure} given by Bowman et al.~\cite{bowman20043d}. We have arranged these terms into three basic categories:

\begin{itemize}[leftmargin=*]
    \item Layout: \textit{representation, structure}
    \item Placement: \textit{location, reference frame}
    \item Interaction: \textit{invocation, removal, highlighting, selection}
\end{itemize}

\subsection{Layout}
Menus in virtual environments come in various shapes and appearances, depending on the use case. Often, menus closely match the scenario's visual appearance, which ensures a consistent experience and benefits the overall game experience~\cite{slater2000virtual}. Other designs preserve a neutral and abstract style, making them familiar and easily recognizable as archetypes of their kind~\cite{dachseltMenuSurv, wegner2017comparison}. Apart from designing the menu itself, research has focused on the menu items' layout and geometry. Over time, many prominent approaches have been proposed, such as the \textit{TULIP} menu~\cite{bowman2001design} or the \textit{Command and Control Cube}~\cite{grosjean2002evaluation}. Many publications have covered the differences between various layouts regarding efficiency and intuitiveness~\cite{santosMenus}. As these approaches mainly have emphasized the fast selection of few distinct menu options, the results are not applicable to inventories aiming to easily manage dozens of items. 

 \begin{figure}[t!]
\centering
\includegraphics[width=\columnwidth]{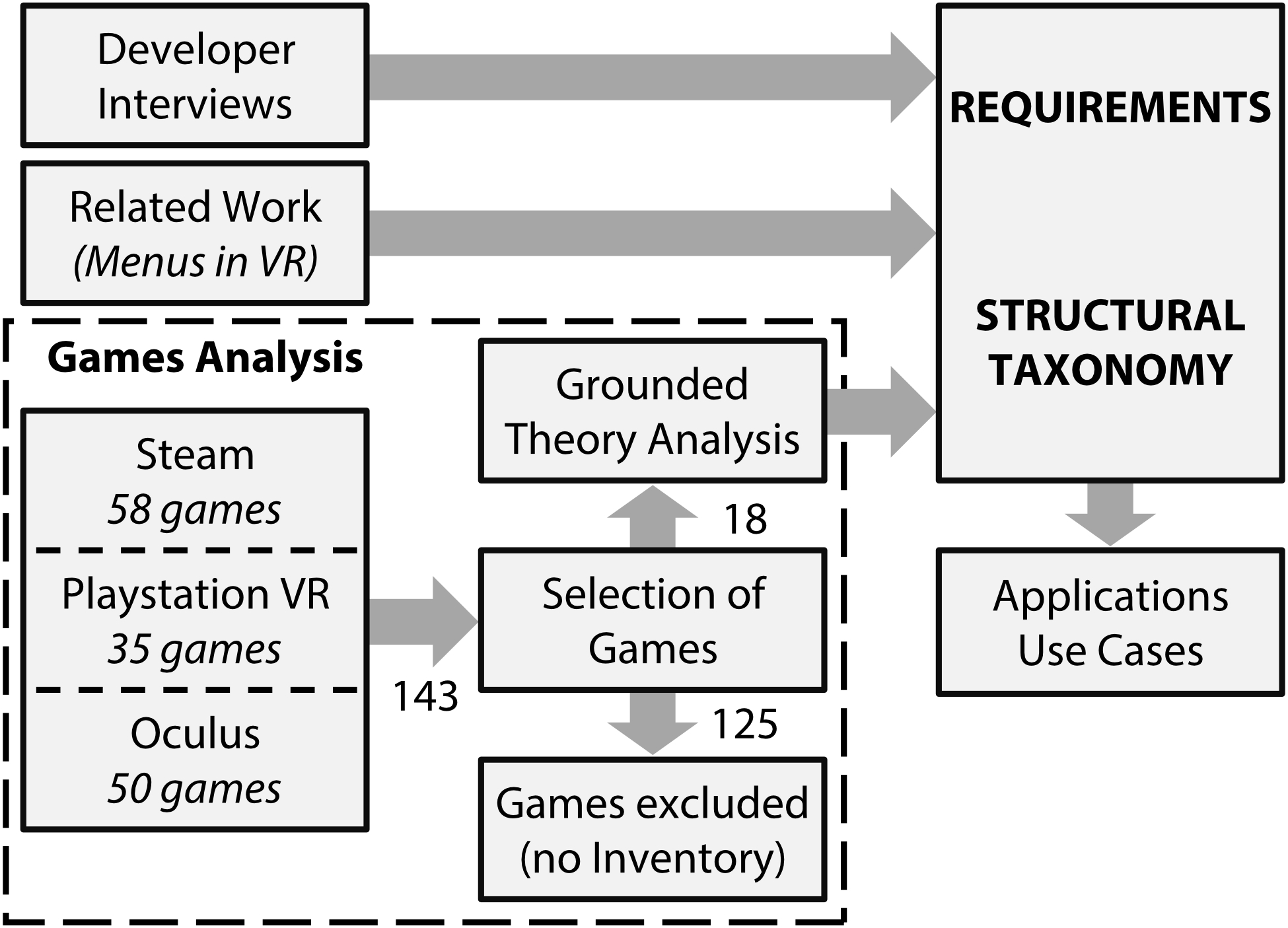}
\caption{Flowchart illustrating our research process, including the related work, developer interviews, and games analysis.}
\label{fig:flowchart}
\end{figure}

\subsection{Placement}
A major challenge when developing VR menus is the third dimension. In contrast to desktop applications, the menu can be positioned freely within the virtual environment. The additional degree of freedom can easily lead to occlusion effects between the world and the interface~\cite{argelaguetSurvSelect, jacobyVirtualMenu}, which are prevented by allowing the players to rearrange the menu at need~\cite{argelaguetSurvSelect}. Before placing the menu into the virtual world, developers must decide on the point of reference. Dachselt et al.~\cite{dachseltMenuSurv} present five possible domains: \textit{world, object, head, body, device}. Past research has emphasized the benefits of bodily interactions: Exploiting the human proprioception could compensate partially for the missing haptic feedback~\cite{lindeman2DInt}. Nevertheless, attaching menus to the player's body bears the risk of exhaustion through constant muscle activity~\cite{argelaguetSurvSelect}. 

Further, menus can be categorized as either \textit{diegetic} or \textit{non-diegetic}~\cite{salomoniDiegetic}. A \textit{diegetic} menu is placed within the scenario and can be used just like any other interactable. This feature offers substantial benefits to presence and player experience, but it usually requires visual embedding and a more realistic reference point, such as the player's body.

\subsection{Interaction}
Interacting with menus comprises three actions: opening, closing, and moving. The first two controls are usually implemented using buttons or simple gestures~\cite{mineIsaac, salomoniDiegetic} whereas moving menus requires at least a three-dimensional (3D) input. Apart from these features, most research has focused on the interaction with the menu items. Choosing items is decomposed into two sequential subtasks: \textit{highlighting} and \textit{selecting}~\cite{jacobyVirtualMenu,kim2000multimodal}. Players position a pointer in 3D space to highlight an item and confirm their selection with a button click. For a comparison of different highlighting and selection techniques, we point interested readers to the work by Argelaguet and Andujar~\cite{argelaguetSurvSelect}.

Many standard VR menus have been taken from desktop applications and modified for use in virtual environments. However, using 1D or 2D interfaces in a 3D surrounding increases the complexity and may induce interaction errors~\cite{handSurvInt}. Also, menus can be quickly out of reach for physical actions~\cite{jacobyVirtualMenu}. Therefore, a popular selection technique is the virtual raycast~\cite{mineIsaac, santosMenus}, which requires only minimal muscle activation~\cite{argelaguetSurvSelect} and can target menus at every distance. An alternative is the virtual hand, with which players use spatially tracked controllers to interact with objects like in the real world. Although slower, more tedious, and limited to the user's range, this approach provides benefits to agency and presence~\cite{welch1996effects}. Many applications use raycasts for menus and virtual hands for all other gameplay interactions. Poupyrev et al.~\cite{poupyrevSelect} did not find differences between the techniques regarding error-rate or selection speed.

\section{Gathering Input from Developers}
A major step towards a comprehensive guideline that can help practitioners in their design process is to ask the community. This decision ensures the practicability of the results and prevents working in an ivory tower. Therefore, we recruited twelve experienced VR game developers from different studios through various VR developer channels, such as Discord, and questioned them using a semi-structured interview. A team member with no experience in the communities held the conversations to avoid any bias through prior contacts. Our primary research questions were:

\noindent \begin{itemize}[leftmargin=*]\compresslist
    \item RQ1: How important are inventory systems for VR games?
    \item RQ2: How difficult is the design and development process?
    \item RQ3: Are the available resources a sufficient aid?
    \item RQ4: What are the unique requirements and challenges when implementing inventories for virtual scenarios?
\end{itemize}

\noindent Even though all participants received the same questions, the interviews mostly followed personal experiences. We analyzed the interview data using a peer-reviewed deductive thematic analysis~\cite{braun2006thematicAnalysis}. The four predetermined main themes followed our initial research questions: Importance/Benefits, Perceived Difficulty, Helpful Resources, and Unique Challenges. 

\noindent\subsection*{\textbf{RQ1:} \textit{How important are inventory systems for VR games?}}

All participants agreed on the general importance of such interfaces for VR games, as they could provide essential benefits to the game and development process. The most frequently mentioned advantage was a more straightforward design phase when adding multiple abilities to a game. Instead of requiring complicated controls, developers could rely on \textit{"a bunch of distinct tools and a common space to store them"(\textbf{D2})}. This approach is especially valuable, considering the limited amount of available buttons on each controller. Another benefit lies within the nature of an inventory: Storing and carrying multiple items enables \textit{"novel gameplay techniques and a deeper storyline"(\textbf{D5})}. Interestingly, one participant reported limiting storage to handheld items. Players are forced to decide carefully what to carry with them and to remember where they left things. This approach is the minimalist version of a fixed capacity inventory not requiring any additional interface.

\noindent\subsection*{\textbf{RQ2:} \textit{How difficult is the design and development process?}}

The second question split the participants into two groups. Half of the developers stated they were actively avoiding the use of inventory systems at all, despite its vast potential. The main reason is the high complexity of the design and implementation phase. Considering the mainly small team sizes, most developers could not afford to spend extraordinary resources on developing such a challenging feature. Even though the other half used inventories in most games, they reported requiring significant development time and multiple iterations since \textit{"it just did not feel natural"(\textbf{D1})}. Overall, the subjects described the topic as being highly complicated and requiring detailed domain knowledge.

\noindent\subsection*{\textbf{RQ3:} \textit{Are the available resources a sufficient aid?}}

Ten of twelve subjects noted that inventories are among the most complex VR techniques that receive subpar attention. Being asked what could aid them in their situation, most developers preferred \textit{"having an off-the-shelf asset to handle it"(\textbf{D9})}. However, the requirements for such a component would be immense since every game requires \textit{"creating a style that looks and feels natural"(\textbf{D11})}. Other commonly requested resources are general guidelines or tutorials on developing inventories that convey a good user experience. Both requests require detailed domain knowledge in a novel research area. We want to address these issues by organizing the design range with precise requirements and a clear structure. 

\noindent\subsection*{\textbf{RQ4:} \textit{What are the unique requirements and challenges when implementing inventories for virtual scenarios?}}

One benefit of virtual setups is the ability to become fully immersed in the scenery with high levels of agency and presence. The participating developers emphasized this unique advantage and underlined the importance of preserving a consistent and natural experience. Inventories should be placed within the virtual world and \textit{"must not appear as an artificial overlay"(\textbf{D12})}. Furthermore, the use of two-dimensional (2D) interfaces is strongly discouraged. Most developers described such inventories as detrimental to the player experience: \textit{"If it is just another flat 2D experience in VR, I feel that will shatter the immersion -- which is what VR is all about"(\textbf{D4})}.\par

Another critical challenge is the positioning of the inventory. The 3D nature of virtual scenarios adds additional difficulty to visibility and usability. Many developers aim for free locomotion within the world, without limiting the accessibility of the inventory: \textit{"Menus and controls must be placed far enough away from the player as to not crowd them, yet close enough to interact"(\textbf{D7})}. One commonly used approach is to attach the inventory directly to the player itself. However, VR players are usually not fully tracked and will not see their own body except the hands. This impediment hinders inventory attachments: \textit{"A belt seems awkward, backpacks seem to stop the game, wrist-based seems the best so far"(\textbf{D9})}.\par

Finally, the developers emphasized using the full range of available controls to convey a realistic and fun experience. Pointing and clicking are well-known interactions that closely resemble traditional computer usage but lack natural counterparts. Instead, subjects prefer fully tracked controllers (6DOF) to implement standard grabbing behavior.  Combining this feature with context-sensitive gestures could achieve even more intuitive and realistic controls: \textit{"A reach over the shoulder is a great system for grabbing a weapon"(\textbf{D8})}.

The overall feedback shows that inventories are an exciting and relevant topic. The various concerns, problems, and challenges faced by VR developers underline the need for sophisticated guidelines. The answers to RQ3 demonstrate that the currently available resources, best practices, and current games are not of sufficient help. Therefore, this work provides a first structural approach to this vital topic.

\begin{table}
\caption{The list of all 18 examined VR games featuring inventories.}%
  \label{tab:gameAnalysisGames}
  \begin{tabular}{p{13mm} p{15mm} p{38mm} r}
    \toprule
    \textbf{Platform} & \textbf{Genre} & \textbf{Game}&\\
    \toprule
    PlayStation & role-playing & The Elder Scrolls V: Skyrim VR & \cite{GameElderScrollsSkyrim}\\
    PlayStation & action &  The Mage's Tale & \cite{GameMagesTale}\\
    PlayStation & adventure &  ARK Park & \cite{GameARKPark}\\
    PlayStation & action &  No Man's Sky VR & \cite{GameNoMansSkyVR}\\
    Steam & role-playing & Crawling Of The Dead & \cite{GameCrawlingDead}\\
    Steam & role-playing & Vanishing Realms & \cite{GameVanishingRealms}\\
    Steam & simulation & Afloat & \cite{GameAfloat}\\
    Steam & shooter & Arizona Sunshine & \cite{GameArizonaSunshine}\\
    Steam & survival & Castaway VR & \cite{GameCastaway}\\
    Steam & survival & Star Shelter & \cite{GameStarShelter}\\
    Steam & survival & The Forest VR & \cite{GameTheForest}\\
    Steam & shooter & Half-Life: Alyx & \cite{Alyx}\\
    Oculus & role-playing & Asgard's Wrath & \cite{AsgardsWrath}\\
    Oculus & survival & Subnautica & \cite{GameSubnautica}\\
    Oculus & shooter & Onward & \cite{GameOnward}\\
    Oculus & shooter & STAND OUT: VR Battle Royale & \cite{GameStandOut}\\
    Oculus & adventure & Batman: Arkham VR & \cite{GameArkhamVR}\\
    Oculus & adventure & The Gallery - Call of the Starseed & \cite{GameGallery}\\
    \bottomrule
  \bottomrule
\end{tabular}
\end{table}

\section{Analyzing VR Games Using Grounded Theory}
After reassuring ourselves of the demand for a general guideline through developer interviews, we conducted a qualitative study on inventories in VR games to identify the essential building blocks and design choices. We used a grounded theory approach adapted from the analysis of idle games by Alharthi et al.~\cite{alharthi2018playing}. \textit{Grounded theory}~\cite{glaser1978theoretical, glaser1998doing, glaser1968discovery} is used to explore novel domains and build a theory from collected data. The approach consists of three steps: In \textit{open coding}, the collected data is structured by applying preliminary labels. The resulting codes are combined into concepts sharing a common theme. These results are further refined in \textit{axial coding} by identifying relationships between the codes and concepts to merge them into categories. Finally, \textit{selective coding} is used to form a general theory using the established categories. Our analysis process, seen in Figure~\ref{fig:flowchart}, encompassed four consecutive steps: games-selection, initial observations, open coding, and axial and selective coding.

\subsection*{Step 1: Selecting VR Games}

We started by identifying VR games that feature inventories. Despite the late popularity, the overall corpus of VR games remains sparse. Many of the available titles are merely demos of non-VR games or experimental micro-games. Additionally, most of the games are distributed on more than one platform. Therefore, we decided to include only distinct games featuring enough content for evaluation and having at least ten reviews in the stores. From the three biggest platforms, \textit{Steam}~\cite{steam}, \textit{Oculus}~\cite{oculus}, and \textit{PlayStation VR}~\cite{psvr}, we chose a total of 143 games. We reviewed all games to determine whether a title was appropriate for our analysis and excluded games with no or minimal inventories. For instance, the game \textit{Moss}~\cite{Moss} featured a menu button to switch between three different styles of the hero's main weapon. Considering this option was purely cosmetic; it did not add any value to the gameplay. In the end, we finished with a corpus of 18 VR-games (see Table~\ref{tab:gameAnalysisGames}).

\begin{table}
  \caption{Exemplary game observations from the first analysis step.} %
  \label{tab:gameAnalysis1}
  \begin{tabular}{l  p{63mm}}
    \toprule
    \textbf{Feature} & \textbf{Observation} \\
    \toprule
    title & Crawling of the Dead~\cite{GameCrawlingDead}\\
    platform & Steam\\
    layout & shape: backpack (fitting game's theme) \newline menu: 2D, floating in front of bag \newline structure: mix of purpose slots \& grid \newline items: miniaturized versions of original object\\
    placement & reference: virtual object floating in the world \newline position: players place inventory freely in front of them \\
    interaction & open/close: use gesture  \newline collect: automatic (touching) \& manual (virtual hand)\\
    notes & sorting partially possible (free slots), except loot\\

    \bottomrule
  \bottomrule
\end{tabular}

\end{table}

\subsection*{Step 2: Observations}
Two researchers went through all 18 games, using trailers, game descriptions, reviews, and gameplay sessions to catch all necessary details of the inventory. For each game, both reviewers completed a predefined table (see Table~\ref{tab:gameAnalysis1}) that followed the general structure of menus (cf. Section~\ref{RelatedWork}).

\subsection*{Step 3: Open Coding}
The results from the previous stage were used to derive the first labels. Using \textit{open coding}, we analyzed the data and generated the first codes describing each inventory's aspects. We unified similar codes into a set of early concepts and categories shared by multiple games. For instance, the code \textit{thematic style} represents an inventory fitting the game's style very closely. Alternatively, other inventories were marked with the code \textit{abstract style} to indicate a neutral user interface that does not reflect the actual theme. The whole process was done by hand in a joint discussion session.

\subsection*{Step 4: Axial and Selective Coding}
\begin{figure}
\centering
\includegraphics[width=0.85\linewidth]{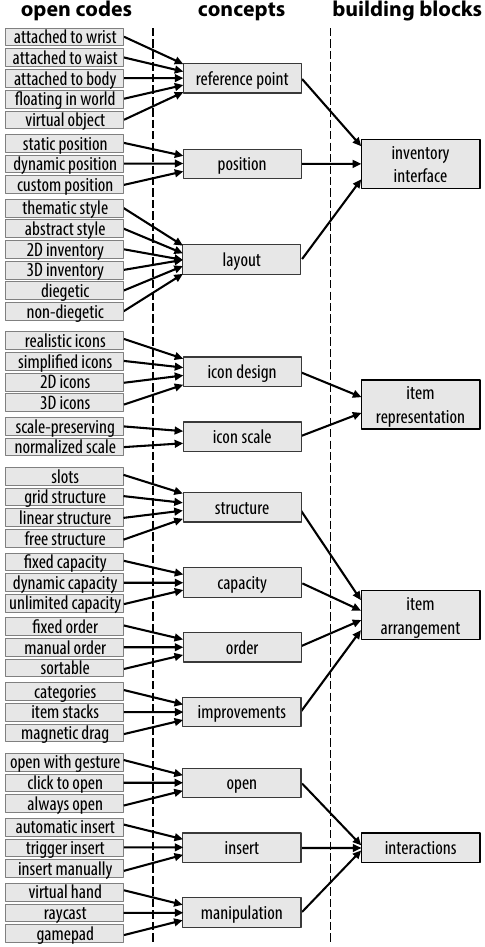}
\caption{Grounded theory analysis: The initial observations are formalized into open codes, which are structured into concepts and overall building blocks.}
\label{fig:groundedTheory}
\end{figure}

The resulting codes, concepts, and early categories were discussed in multiple sessions to combine them into logical units. We reassigned the games to new codes, consulted related work, and revisited the games to achieve a universal set of categories named \textit{building blocks}. Each block contained two to four concepts with underlying codes. For instance, all inventories either \textit{preserved} the item's scale or \textit{normalized} the size to fit the structure. These two codes describe the item's \textit{scale} within the inventory and form a concept as part of the general building block \textit{item representation}. The three others are \textit{interface}, \textit{item arrangement}, and \textit{interactions}. Figure~\ref{fig:groundedTheory} depicts the complete process.

\section{Results}
\begin{figure*}[t!]
\centering
\includegraphics[width=2.045\columnwidth]{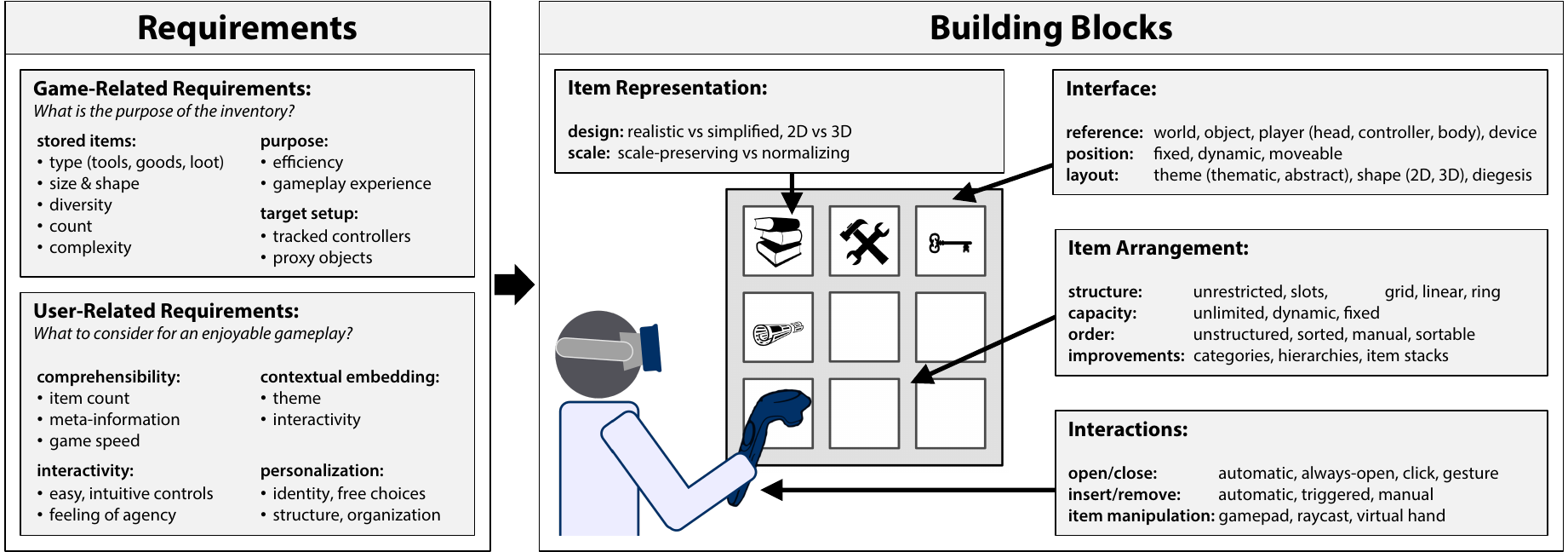}
\caption{Requirements and taxonomy of inventory systems in virtual environments. This figure is read from the left, starting with the requirements that should be taken into account before designing inventories. The considerations are used to select the design choices in the taxonomy on the right.}
\label{fig:taxonomy}
\end{figure*}

Based on the related work, developer feedback, and games analysis, this section summarizes the main characteristics of inventory systems in virtual environments. First, we assess the game- and user-related requirements that need to be taken into account when designing an inventory for a particular use case. Afterward, we decompose the structure of inventories into a universal taxonomy by explaining the different building blocks and identified design choices (see Figure~\ref{fig:taxonomy}).

\subsection{Game-Related Requirements}
Using an inventory can provide significant benefits to many games. The success of a particular implementation depends on the interplay between the game and the storage feature. Therefore, every design phase should begin with a careful analysis of the use case and the required features.

The key factor determining the design process is the stored item. Usually, items are subdivided into three groups: \textit{"tools, goods, and loot"(\textbf{D2})}. Tools and goods are frequently utilized, whereas loot is mostly used for acquiring wealth. This subdivision leads to two primary inventory types: \textit{Carry} inventories focus on few, quickly accessible items, e.g., in the game \textit{Fortnite}~\cite{GameFortnite}. \textit{Diablo III}~\cite{GameDiablo3} is a perfect remedy for a \textit{loot} inventory, providing easily manageable storage~\cite{hamari2010game, Sztajer2010Inventory}. Furthermore, considering the complexity and variety of included items is essential. Games relying on few, diverse objects need different designs than games with large amounts of similar items. 

Apart from item considerations, the inventory's particular purpose is decisive for any design. Fast-paced action games require an efficient layout limiting the features in favor of speed, whereas RPGs may interweave a more complex design with the gameplay, e.g., through \textit{"managing limited storage spaces"(\textbf{D12})}. The game's target platforms determine the available capabilities: Early mobile HMDs were constrained to rotational controllers, whereas current setups provide positional tracking or the ability to track physical proxies.

\subsection{User-Related Requirements}
One reason for failed inventory designs is poor fitting to the game's characteristics. Many interviewed developers stated that their implementations often \textit{"just did not feel natural"(\textbf{D1})}. This feedback illustrates that designers should focus not only on the structural requirements but also the player experience. The proposed design considerations in this section can help determine whether a particular design fits the special requirements of VR and conveys a proper user experience.

\subsubsection{\textbf{Comprehensibility}} The information for every stored item must be displayed in a comprehensible manner. Displaying too many items or meta-information on the limited VR screen can easily lead to visual cluttering~\cite{jacobyVirtualMenu, santosMenus}, which increases the mental effort and can spoil the whole gameplay~\cite{Sztajer2010Inventory}.

\subsubsection{\textbf{Interactivity}} Managing and interacting with the inventory and the stored items should be as easy and quick as possible~\cite{argelaguetSurvSelect}. For instance, situation-dependent controls may simplify the necessary user actions to a minimum. However, such game designs bear the risk of reducing the feeling of agency~\cite{Sztajer2010Inventory}. Especially in VR, players like to interact with the environment and to feel in control over the resulting actions. \par

\subsubsection{\textbf{Contextual Embedding}} Even the best inventory design will not be well received without matching the enclosing game. An interactive system should always fit the provided context regarding theme and interactivity. A linear story-driven game does not need a fully fetched inventory with categories and sorting methods. Instead, a slot-based storage integrated within the game's theme fits much better to the limited player abilities.

\subsubsection{\textbf{Personalization}} Inventories as personal and individual spaces have a considerable impact on the game experience. Especially in linear games, players have little to no chance to individualize their gameplay. The exception is an inventory, which provides complete freedom over content and organization~\cite{Sztajer2010Inventory}. Thus, freely manageable inventories can provide significant advantages to character identification and presence.

\subsection{Structural Taxonomy}
Next, we use the identified concepts from the previous analysis-step to disassemble the inventory into its components. We explain each building block and propose design recommendations based on related research and developer feedback.

\subsubsection{\textbf{Interface}}
The interface is the fundamental component containing all storage items and determining the inventory's position, shape, and design. During the invocation, the interface is bound to a reference point used as an initial positional anchor. Following the definition given in the related work, possible reference points are world, object, head, body, controller, and device. The chosen reference determines the inventory's initial location. Fixed inventories retain this position until disposal. Alternatively, dynamic interfaces are attached to the reference point and follow all positional changes. Both approaches might cause occlusion effects between surrounding and inventory, which could easily lead to frustration and decreased usability. Therefore, it might be better to give the player the chance to move the storage freely. This feature could also provide further benefits in terms of personalization and interactivity.

Deciding on an interface layout, developers can choose between a 2D or a 3D style. Most analyzed games use a 2D design, which simplifies the development process by using established interactions. However, those interfaces were collectively rejected by all interviewed developers as they are believed to \textit{"severely ruin the immersion and defeat the purpose of VR"(\textbf{D2})}. The alternative is a 3D interface providing the basis for more realistic and better integrated implementations.

Apart from the shape, the layout is also determined by the thematic and diegetic fitting. Usually, VR games aim to maximize immersion into the virtual scenery and avoid any thematic cuts. Therefore, matching the inventory's style as closely as possible to the surrounding environment is natural. In contrast, abstract menus offer the advantage of prior knowledge: Most people have already experienced similar storage interfaces and are proficient to a certain degree. Therefore, an abstract design helps to reduce the necessary cognitive load. Diegetic interfaces maximize thematic embedding and become a part of the game world, e.g., as a backpack~\cite{GameCrawlingDead, Alyx}. A fully diegetic inventory immerses completely into the environment and removes any perceivable cut reducing the player's presence.

\subsubsection{\textbf{Item Representation}}
Every stored item needs a graphical or textual representation. The chosen concept strongly influences the amount of conveyable information. Many games need to provide more data than the basic item's appearance. Some of the most common item information are category, count, usability, or value. However, the available display area forces developers to reduce the information to a minimum and rely on meaningful representations to convey the details space-effectively. Realistic designs allow for detailed conclusions on the object's shape and physical properties while merging into the virtual scenario. In return, a more simplified style, such as icons or texts, reduces the visual clutter and allows for increased information density. The choice between a 2D and 3D representation is usually based on the interface type and not directly linked to the degree of realism.

One key aspect of the chosen representation is the displayed size: Preserving the object's original scale removes any discrepancy between the item and its representation in the inventory. However, this design choice could quickly fill the limited space with large items and introduce occlusion problems. The alternative is to normalize the item's scale upon insertion.

\subsubsection{\textbf{Item Arrangement}}
According to the interviewed developers, the item arrangement is at least as vital as the representation. Depending on the use case, the layout should focus on either accessibility, management, or overview of the content. Most of the analyzed VR games use grids or fixed purpose slots. Few games grant more freedom to the player and support free object placement. Despite providing benefits for presence and personalization, this decision bears the risk of producing obstructive and chaotic inventories. Apart from the item arrangement, other reasons for poor comprehensibility are massive storage sizes or extreme information densities. The typical solutions are limiting the inventory capacity or reducing the provided information per item. Many games mitigate this problem by extending the maximum storage during the journey.

Several inventories provide features to sort the stored items. These range from complete flexibility and self-administered organization, through optional sorting techniques, to fixed automatic orders. After all, inventories should maximize the player's freedom and control over the item arrangement without sacrificing too much comprehensibility. The overview can be improved by including concepts such as categories, hierarchies, or item stacks. Some games limit the inventory use to tools and automate the collection and management of loot.

\begin{figure}[t!]
\centering
\includegraphics[width=\columnwidth]{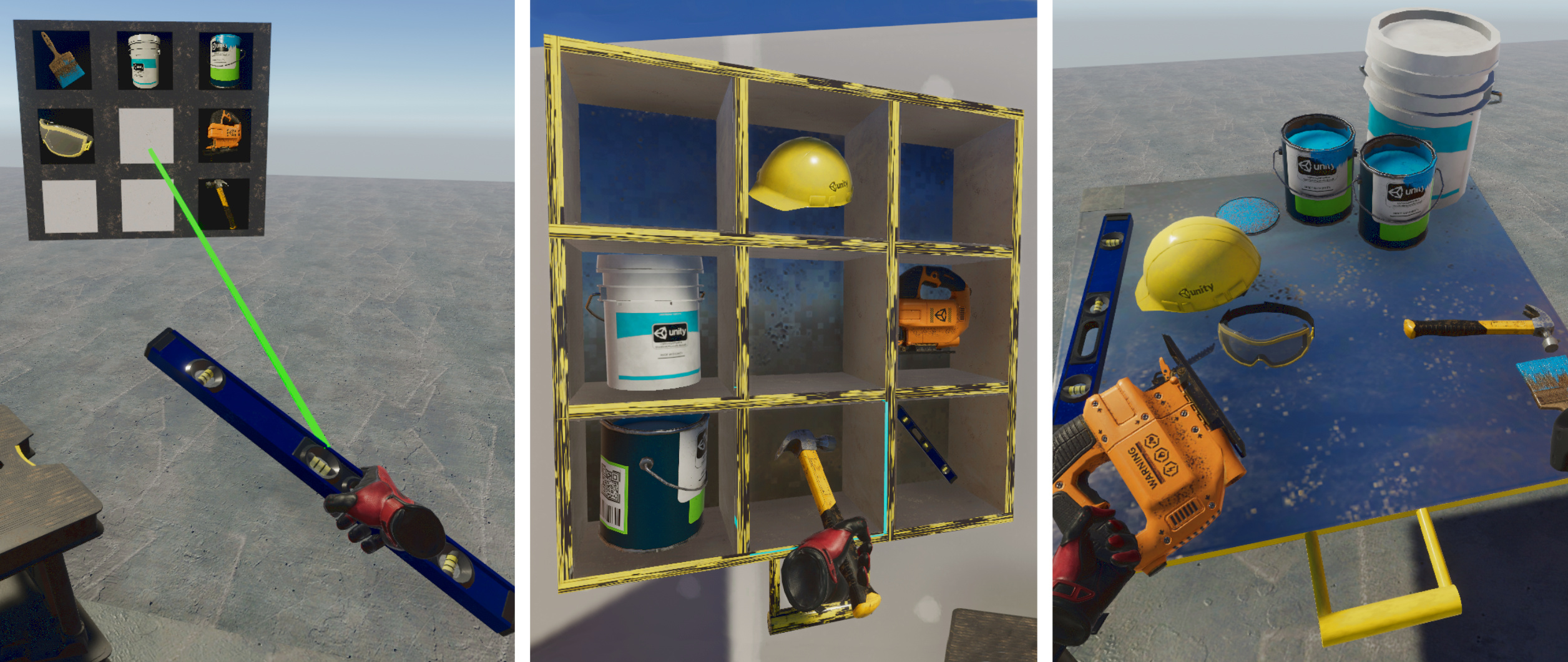}
\caption{Our three inventory prototypes: \textit{Flat Grid} (left), \textit{Virtual Drawers} (middle), and \textit{Magnetic Surface} (right).}
\label{fig:invTogether}
\end{figure}

\subsubsection{\textbf{Interactions}}
The novel VR interaction concepts made possible through spatially tracked controllers are by far the biggest difference compared to non-VR games. Instead of using a simple button click, games could implement more natural ways to open the inventory. An example is \textit{The Gallery}~\cite{GameGallery}, where players reach behind themselves to retrieve a virtual backpack. Interestingly, many examined games do not support opening the inventory but keep it visible. The second major inventory interaction is the item insertion. We identified three approaches: Games require the players to insert items manually or feature automatic or button-triggered collection.

Every inventory requires an interaction technique for item manipulation. The various approaches include classical gamepad controls, raycast-aiming, and physical actions through virtual hands. Direct interactions with tracked controllers require the least learning time and outperform distant object manipulation significantly~\cite{mine1997moving}. Additionally, physical grabbing avoids typical problems of virtual pointers, e.g., Fitt's law~\cite{fitts1954information}, and is deemed the gold standard for interactive gameplay. Our results reflect this trend towards more intuitive controls. Only two VR adaptations still rely on the gamepad~\cite{GameElderScrollsSkyrim, GameSubnautica}. A minority of five games use the raycast-aiming, whereas the others prefer the virtual hand. Despite being the most preferred technique, physical grabbing is usually not applicable for games relying heavily on distant object manipulation. The game \textit{The Mage's Tale}~\cite{GameMagesTale} tackles this problem creatively by providing a magnetic force ability to drag any item into the own hand.

\section{Designing Inventories}
In this section, we assess the practical applicability of our taxonomy by developing three inventories for different use cases (see Figure~\ref{fig:invTogether}) and explaining the underlying design considerations. We assume that these insights will help practitioners with their design process. As designing an inventory depends largely upon the intended use case, we use our designs to target three different exemplary goals:

\begin{enumerate}[leftmargin=*]\compresslist
    \item simplicity and performance 
    \item natural and intuitive interactions
    \item engaging and interactive gameplay
\end{enumerate}

\subsection{Flat Grid -- A "Swiss Army Knife"}
The prototype aims to be a simplistic and universal interface. It focuses on fast interaction and low visual complexity, useful for fast-paced games requiring efficient item management, e.g., of loot. These demands call for a more abstract and well-known design. As players access it only briefly to place or retrieve an item, it does not need an option to rearrange the positioning. Following these considerations, the interface appears on command at a fixed position in front of the player. An abstract 2D overlay arranging items in a regular grid reduces the visual clutter and makes it easy to scan the contents. Also, the stored items are reduced to simple 2D icons of equal sizes. Finally, we maximize the interaction speed by using a virtual raycast pointer to interact with the stored items.

\subsection{Virtual Drawers -- Featuring Natural Interaction}
The second design focuses on natural interactions while preserving a clear storage structure. These interactions reduce the necessary learning time and the possibility of misuse, but they usually involve a considerate amount of physical effort and are less suited for huge arsenals of items. Recent VR games have undergone a general shift towards more realistic handling of few meaningful objects. For instance, players use a single upgradable weapon instead of collecting dozens of different arms. Such games can profit from a more natural and intuitive inventory where performance is negligible. Compared to the \textit{Flat Grid}, this use case requires a more realistic interface: A diegetic 3D shelf fits the scenario while preserving the same organized structure. As the original 2D icons no longer fit the 3D interface, we instead scale the stored item to a uniform size. This solution preserves the original shape while accounting for individual size differences. Finally, the design focuses on natural interactions instead of speed. Therefore, the items are stored in the inventory by placing them physically into one of the free slots. The players may also grab the whole inventory using a handle and move it freely within the environment.

\subsection{Magnetic Surface -- Inventories as Gameplay Element}
The final design demonstrates the use of inventories as a core gameplay element. In games like \textit{Green Hell}~\cite{GameGreenHell}, players spend much of their playtime arranging the items within their backpack to counter limited carrying capacity. Instead of focussing on performance, such designs aim for innovative experiences. Usually, there is no generic approach as such implementations are highly application specific. For our particular use case, we focus on a novel VR experience offering maximal freedom. Unlike the previous design, this prototype preserves the item's shape and scale, allowing players to distinguish the items based on their shapes and sizes. We replaced the grid with a simple rectangular worktop floating in the air. Players can put any object onto the plate where magnetic force keeps the items in place. Consequently, players are in full control over item arrangement and positioning without any form of forced organization. The inventory does not limit the item density, which provides a dynamic capacity based on the size of the worktop and the player's abilities.

\section{Conclusion}

Developers of VR inventories can select from various designs for each component and thus must consider the framing determined by the respective game. Our work introduced this timely topic by structuring the research area in multiple steps. After filtering the applicable, related work, we conducted semi-structured interviews with developers to assess the community's needs. Then, we analyzed the inventories of 18 VR games and decomposed the interfaces' structure. Our taxonomy covers the vast design space of VR inventories, ranging from simple 2D solutions to diegetic interfaces. While these building blocks share common aspects with non-VR inventories, they also account for VR-related peculiarities, such as spatial interactions. Structuring the design process into requirements and building blocks provides a guideline and facilitates decision-making. We emphasize that our work is not limited to real-world inventories but also applies to storage concepts beyond those known from our daily lives. In the final part, we designed three inventories for specific use cases to demonstrate our contribution. 

The application section also unveils our open questions: Developers usually consider game- and user-related requirements first. These assumptions provide an idea of the critical aspects required for the next implementation steps. However, how these considerations lead to specific design decisions favorable for the intended use case remains unclear. Next, developers must assemble a complete set of design choices. This selection bears further challenges in the form of detrimental effects between individual design elements. These problems illustrate the need to evaluate the interplay between requirements and design choices and the mutual effects between building blocks further. We assume that such future work will complement our structural approach and help developers and researchers.

\bibliographystyle{IEEEtran}
\bibliography{IEEEabrv,literature}

\begin{thebibliography}{10}
\providecommand{\url}[1]{#1}
\csname url@samestyle\endcsname
\providecommand{\newblock}{\relax}
\providecommand{\bibinfo}[2]{#2}
\providecommand{\BIBentrySTDinterwordspacing}{\spaceskip=0pt\relax}
\providecommand{\BIBentryALTinterwordstretchfactor}{4}
\providecommand{\BIBentryALTinterwordspacing}{\spaceskip=\fontdimen2\font plus
\BIBentryALTinterwordstretchfactor\fontdimen3\font minus
  \fontdimen4\font\relax}
\providecommand{\BIBforeignlanguage}[2]{{%
\expandafter\ifx\csname l@#1\endcsname\relax
\typeout{** WARNING: IEEEtran.bst: No hyphenation pattern has been}%
\typeout{** loaded for the language `#1'. Using the pattern for}%
\typeout{** the default language instead.}%
\else
\language=\csname l@#1\endcsname
\fi
#2}}
\providecommand{\BIBdecl}{\relax}
\BIBdecl

\bibitem{GameGreenHell}
{Creepy Jar}, ``\emph{Green Hell},'' Game [PC], August 2018.

\bibitem{handSurvInt}
C.~Hand, ``A survey of 3d interaction techniques,'' in \emph{Computer graphics
  forum}, vol.~16, no.~5.\hskip 1em plus 0.5em minus 0.4em\relax Wiley Online
  Library, 1997, pp. 269--281.

\bibitem{wegner2017comparison}
K.~Wegner, S.~Seele, H.~Buhler, S.~Misztal, R.~Herpers, and J.~Schild,
  ``Comparison of two inventory design concepts in a collaborative virtual
  reality serious game,'' in \emph{Extended Abstracts Publication of the Annual
  Symposium on Computer-Human Interaction in Play}.\hskip 1em plus 0.5em minus
  0.4em\relax ACM, 2017.

\bibitem{cmentowski2019inventoryLBW}
S.~Cmentowski, A.~Krekhov, A.-M. M\"{u}ller, and J.~Kr\"{u}ger, ``Toward a
  taxonomy of inventory systems for virtual reality games,'' in \emph{Extended
  Abstracts of the Annual Symposium on Computer-Human Interaction in Play
  Companion Extended Abstracts}.\hskip 1em plus 0.5em minus 0.4em\relax ACM,
  2019, pp. 363--370.

\bibitem{dachseltMenuSurv}
R.~Dachselt and A.~H{\"u}bner, ``Three-dimensional menus: A survey and
  taxonomy,'' \emph{Computers \& Graphics}, vol.~31, no.~1, pp. 53--65, 2007.

\bibitem{kim2000multimodal}
N.~Kim, G.~J. Kim, C.-M. Park, I.~Lee, and S.~H. Lim, ``Multimodal menu
  presentation and selection in immersive virtual environments.'' in \emph{vr},
  2000, p. 281.

\bibitem{bowman20043d}
D.~Bowman, E.~Kruijff, J.~J. LaViola~Jr, and I.~P. Poupyrev, \emph{3D User
  interfaces: theory and practice, CourseSmart eTextbook}, 2004.

\bibitem{jacobyVirtualMenu}
R.~H. Jacoby and S.~R. Ellis, ``Using virtual menus in a virtual environment,''
  in \emph{Visual Data Interpretation}, vol. 1668.\hskip 1em plus 0.5em minus
  0.4em\relax International Society for Optics and Photonics, 1992, pp. 39--48.

\bibitem{slater2000virtual}
M.~Slater and A.~Steed, ``A virtual presence counter,'' \emph{Presence:
  Teleoperators \& Virtual Environments}, vol.~9, no.~5, pp. 413--434, 2000.

\bibitem{bowman2001design}
D.~A. Bowman and C.~A. Wingrave, ``Design and evaluation of menu systems for
  immersive virtual environments,'' in \emph{Proceedings IEEE Virtual Reality
  2001}.\hskip 1em plus 0.5em minus 0.4em\relax IEEE, 2001, pp. 149--156.

\bibitem{grosjean2002evaluation}
J.~Grosjean, J.-M. Burkhardt, S.~Coquillart, and P.~Richard, ``Evaluation of
  the command and control cube,'' in \emph{Proceedings of the 4th IEEE
  international conference on multimodal interfaces}, 2002, p. 473.

\bibitem{santosMenus}
A.~Santos, T.~Zarraonandia, P.~D{\'\i}az, and I.~Aedo, ``A comparative study of
  menus in virtual reality environments,'' in \emph{Proceedings of the 2017 ACM
  International Conference on Interactive Surfaces and Spaces}, 2017.

\bibitem{argelaguetSurvSelect}
F.~Argelaguet and C.~Andujar, ``A survey of 3d object selection techniques for
  virtual environments,'' \emph{Computers \& Graphics}, vol.~37, 2013.

\bibitem{lindeman2DInt}
R.~W. Lindeman, J.~L. Sibert, and J.~K. Hahn, ``Hand-held windows: towards
  effective 2d interaction in immersive virtual environments,'' in
  \emph{Proceedings IEEE Virtual Reality (Cat. No. 99CB36316)}.\hskip 1em plus
  0.5em minus 0.4em\relax IEEE, 1999.

\bibitem{salomoniDiegetic}
P.~Salomoni, C.~Prandi, M.~Roccetti, L.~Casanova, and L.~Marchetti, ``Assessing
  the efficacy of a diegetic game interface with oculus rift,'' in \emph{2016
  13th IEEE Annual Consumer Communications \& Networking Conference
  (CCNC)}.\hskip 1em plus 0.5em minus 0.4em\relax IEEE, 2016, pp. 387--392.

\bibitem{mineIsaac}
M.~Mine, ``Isaac: A virtual environment tool for the interactive construction
  of virtual worlds,'' \emph{UNC Chapel Hill Computer Science Technical Report
  TR95-020}, 1995.

\bibitem{welch1996effects}
R.~B. Welch, T.~T. Blackmon, A.~Liu, B.~A. Mellers, and L.~W. Stark, ``The
  effects of pictorial realism, delay of visual feedback, and observer
  interactivity on the subjective sense of presence,'' \emph{Presence:
  Teleoperators \& Virtual Environments}, vol.~5, no.~3, pp. 263--273, 1996.

\bibitem{poupyrevSelect}
I.~Poupyrev, S.~Weghorst, M.~Billinghurst, and T.~Ichikawa, ``A study of
  techniques for selecting and positioning objects in immersive ves: effects of
  distance, size, and visual feedback,'' in \emph{Proceedings of ACM CHI},
  vol.~98, 1998.

\bibitem{braun2006thematicAnalysis}
V.~Braun and V.~Clarke, ``Using thematic analysis in psychology,''
  \emph{Qualitative research in psychology}, vol.~3, no.~2, pp. 77--101, 2006.

\bibitem{GameElderScrollsSkyrim}
{Bethesda Studios}, ``\emph{The Elder Scrolls V: Skyrim VR},'' Game
  [PlayStation VR], November 2017.

\bibitem{GameMagesTale}
{inXile Entertainment}, ``\emph{The Mage's Tale},'' Game [PlayStation VR],
  2019.

\bibitem{GameARKPark}
{Snail Games}, ``\emph{ARK Park},'' Game [PlayStation VR], March 2018.

\bibitem{GameNoMansSkyVR}
{Hello Games}, ``\emph{No Man's Sky},'' Game [PlayStation VR], August 2016.

\bibitem{GameCrawlingDead}
{Running Pillow}, ``\emph{Crawling Of The Dead},'' Game [SteamVR], 2019.

\bibitem{GameVanishingRealms}
{Indimo Labs LLC}, ``\emph{Vanishing Realms},'' Game [SteamVR], August 2019.

\bibitem{GameAfloat}
{Blue Atom Interactive}, ``\emph{Afloat},'' Game [SteamVR], August 2019.

\bibitem{GameArizonaSunshine}
{Vertigo Games}, ``\emph{Arizona Sunshine},'' Game [SteamVR], December 2016.

\bibitem{GameCastaway}
{MC Games}, ``\emph{Castaway VR},'' Game [SteamVR], October 2018.

\bibitem{GameStarShelter}
{Overflow}, ``\emph{Star Shelter},'' Game [SteamVR], October 2017.

\bibitem{GameTheForest}
{Endnight Games Ltd}, ``\emph{The Forest},'' Game [SteamVR], April 2018.

\bibitem{Alyx}
{Valve}, ``\emph{Half-Life: Alyx},'' VR Game [Windows], March 2020.

\bibitem{AsgardsWrath}
{Sanzaru Games}, ``\emph{Asgard's Wrath},'' VR Game [Oculus], October 2019.

\bibitem{GameSubnautica}
{Unknown Worlds}, ``\emph{Subnautica},'' Game [Oculus], March 2016.

\bibitem{GameOnward}
{Downpour Interactive}, ``\emph{Onward},'' Game [Oculus], November 2017.

\bibitem{GameStandOut}
{Raptor Lab}, ``\emph{STAND OUT: VR Battle Royale},'' Game [Oculus], 2019.

\bibitem{GameArkhamVR}
{Rocksteady Studios}, ``\emph{Batman: Arkham VR},'' Game [Oculus], April 2017.

\bibitem{GameGallery}
{Cloudhead Games Ltd.}, ``\emph{The Gallery - Episode 1: Call of the
  Starseed},'' Game [Oculus], December 2016.

\bibitem{alharthi2018playing}
S.~A. Alharthi, O.~Alsaedi, Z.~O. Toups, J.~Tanenbaum, and J.~Hammer, ``Playing
  to wait: A taxonomy of idle games,'' in \emph{Proceedings of the 2018 CHI
  Conference on Human Factors in Computing Systems}.\hskip 1em plus 0.5em minus
  0.4em\relax ACM, 2018.

\bibitem{glaser1978theoretical}
B.~Glaser, ``Theoretical sensitivity,'' \emph{Advances in the methodology of
  grounded theory}, 1978.

\bibitem{glaser1998doing}
B.~G. Glaser, \emph{Doing grounded theory: Issues and discussions}.\hskip 1em
  plus 0.5em minus 0.4em\relax Sociology Press, 1998.

\bibitem{glaser1968discovery}
B.~G. Glaser, A.~L. Strauss, and E.~Strutzel, ``The discovery of grounded
  theory; strategies for qualitative research,'' \emph{Nursing research}, 1968.

\bibitem{steam}
Valve, ``{Steam},'' Website, 2019, retrieved September 20, 2019 from
  \url{https://store.steampowered.com/}.

\bibitem{oculus}
{Facebook Technologies}, ``{Oculus Rift Store},'' Website, 2019, retrieved
  September 20, 2019 from \url{https://www.oculus.com/experiences/rift/}.

\bibitem{psvr}
Sony, ``{PlayStation Store VR Catalog},'' Website, 2019, retrieved September
  20, 2019 from \url{www.playstation.com/ps-vr/ps-vr-games/}.

\bibitem{Moss}
{Polyarc}, ``\emph{Moss},'' VR Game [Oculus], February 2018.

\bibitem{GameFortnite}
{Epic Games} and {People Can Fly}, ``\emph{Fortnite},'' Game [PC], July 2017.

\bibitem{GameDiablo3}
{Blizzard Entertainment}, ``\emph{Diablo 3},'' Game [PC], May 2012.

\bibitem{hamari2010game}
J.~Hamari and V.~Lehdonvirta, ``Game design as marketing: How game mechanics
  create demand for virtual goods,'' \emph{International Journal of Business
  Science \& Applied Management}, vol.~5, no.~1, pp. 14--29, 2010.

\bibitem{Sztajer2010Inventory}
\BIBentryALTinterwordspacing
P.~Sztajer. Mechanical breakdown: The inventory. [Online]. Available:
  \url{https://kotaku.com/mechanical-breakdown-the-inventory-5612149}
\BIBentrySTDinterwordspacing

\bibitem{mine1997moving}
M.~R. Mine, F.~P. Brooks~Jr, and C.~H. Sequin, ``Moving objects in space:
  exploiting proprioception in virtual-environment interaction.'' in
  \emph{SIGGRAPH}, vol.~97, 1997, pp. 19--26.

\bibitem{fitts1954information}
P.~M. Fitts, ``The information capacity of the human motor system in
  controlling the amplitude of movement.'' \emph{Journal of experimental
  psychology}, vol.~47, no.~6, p. 381, 1954.

\end{thebibliography}

\end{document}